\documentclass[aps,prb,twocolumn,superscriptaddress,showpacs,floatfix]{revtex4-1}
\usepackage{amsmath,amssymb,graphicx}
\usepackage{wasysym}
\usepackage[utf8]{inputenc}
\usepackage[T1]{fontenc}
\usepackage{xcolor}
\IfFileExists{newtxtext.sty}
   {\usepackage{newtxtext,newtxmath}}
   {\IfFileExists{stix.sty}
      {\usepackage{stix}}
      {\IfFileExists{mathptmx.sty}
      {\usepackage{mathptmx}}{} } }
\usepackage{textcomp}

\usepackage{bm}

\IfFileExists{siunitx.sty}{\usepackage{booktabs,siunitx}}{}

\pdfoutput=1
\usepackage{color}
\definecolor{LinkColor}{rgb}{0.256,0.439,0.588}

\usepackage{pifont}

\usepackage{hyperref}
\hypersetup{
	colorlinks = true,
	linkcolor = [rgb]{0.70,0.13,0.13},
	citecolor = [rgb]{0.13,0.55,0.13},
	urlcolor = [rgb]{0.25,0.41,0.88}}

\DeclareSymbolFont{sfletters}{OML}{cmbrm}{m}{it}
\DeclareMathSymbol{\sfeps}{\mathord}{sfletters}{"22}

\graphicspath{{Figures/}}

\begin{document}
\title{Classical model emerges in quantum entanglement: \\Quantum Monte Carlo study for an Ising-Heisenberg bilayer}

\author{Siying Wu}
\thanks{These authors contributed equally to this work}
\affiliation{State Key Laboratory of Surface Physics and Department of Physics, Fudan University, Shanghai 200438, China}
\affiliation{Beihang Hangzhou Innovation Institute Yuhang, Hangzhou 310023, China}

\author{Xiaoxue Ran}
\thanks{These authors contributed equally to this work}
\affiliation{Department of Physics and HKU-UCAS Joint Institute of Theoretical and Computational Physics, The University of Hong Kong, Pokfulam Road, Hong Kong SAR, China}

\author{Binbin Yin}
\affiliation{Department of Architecture and Civil Engineering, City University of Hong Kong, Kowloon, Hong Kong, China}

\author{Qi-Fang Li}
\affiliation{The Institute for Solid State Physics, The University of Tokyo, Chiba 277-8581, Japan}

\author{Bin-Bin Mao}
\email{maobinbin@uor.edu.cn}
\affiliation{School of Foundational Education, University of Health and Rehabilitation Sciences, Qingdao 266071, China}

\author{Yan-Cheng Wang}
\email{wangyancheng@zfau.cn}
\affiliation{Beihang Hangzhou Innovation Institute Yuhang, Hangzhou 310023, China}

\author{Zheng Yan}
\email{zhengyan@westlake.edu.cn}
\affiliation{Department of Physics, School of Science, Westlake University, 600 Dunyu Road, Hangzhou 310030, Zhejiang Province, China}
\affiliation {Institute of Natural Sciences, Westlake Institute for Advanced Study, 18 Shilongshan Road, Hangzhou 310024, Zhejiang Province, China}
\affiliation{Department of Physics and HKU-UCAS Joint Institute of Theoretical and Computational Physics, The University of Hong Kong, Pokfulam Road, Hong Kong SAR, China}
\begin{abstract}
By developing a cluster sampling of stochastic series expansion quantum Monte Carlo method, we investigate a spin-$1/2$ model on a bilayer square lattice with intra-layer ferromagnetic (FM) Ising coupling and inter-layer antiferromagnetic(AFM) Heisenberg interaction. The continuous quantum phase transition which occurs at $g_c=3.045(2)$ between the FM Ising phase and the dimerized phase is studied via large scale simulations. From the analyzes of critical exponents we show that this phase transition belongs to the (2+1)-dimensional (3D) Ising universality class. Besides, the quantum entanglement is strong between the two layers, especially in dimerized phase. The effective Hamiltonian of single layer seems like a transverse field Ising model. However, we found the quantum entanglement Hamiltonian is a pure classical Ising model without any quantum fluctuations. Furthermore, we give a more general explanation about how a classical entanglement Hamiltonian emerges.
\end{abstract}

\date{\today}
\maketitle

\section{Introduction}
Quantum entanglement is a powerful tool to detect and characterize the informational, field-theoretical, and topological properties of quantum many-body states~\cite{vidal2003entanglement,Korepin2004universality,Kitaev2006,Levin2006}, which combines the conformal field theory (CFT) and the categorial description of the problem~\cite{Calabrese2008entangle,Fradkin2006entangle,Nussinov2006,Nussinov2009,CASINI2007,JiPRR2019,ji2019categorical,kong2020algebraic,XCWu2020,XCWu2021,JRZhao2021,JRZhao2022,YCWang2021DQCdisorder}. Since Li and Haldane pointed out that the entanglement spectrum (ES) is an important measurement with more information than the entanglement entropy (EE)~\cite{Li2008entangle,Thomale2010nonlocal,Poilblanc2010entanglement}, low-lying ES has been widely employed as a fingerprint of CFT and topology~\cite{Pollmann2010entangle,Fidkowski2010,Yao2010,Canovi2014,LuitzPRL2014,LuitzPRB2014,LuitzIOP2014,Chung2014,Pichler2016,Cirac2011,Stoj2020,guo2021entanglement,Grover2013,Assaad2014,Assaad2015,Parisen2018,yu2021conformal}. The entanglement Hamiltonian (EH) $\mathcal{H}_A=-{\rm ln}(\rho_A)$ can be obtained from the reduced density matrix (RDM) $\rho_A={\rm Tr}_B (\rho)$ of the system $A$ by tracing out the environment $B$~\cite{Li2008entangle,kokail2021entanglement}. The energy spectra of the entanglement Hamiltonian $\mathcal{H}_A$, which is usually called the entanglement spectrum, describes the rich information of the system $A$.

Most of the numerical studies of ES have focused on (quasi-)1D systems so far, because of the limitations of the numerical methods. For example, due to the exponentially growth of the computational complexity and memory cost, the numerical methods such as exact diagonalization (ED) and density matrix renormalization group (DMRG) have therefore obvious limitations for entangling region with long boundaries of lattice and higher dimensions. Recently, a novel scheme has been proposed to extract the information of high dimensional entanglement spectrum by using quantum Monte Carlo (QMC), which opens a new way to study the ES and EH of many-body systems~\cite{2021arXiv211205886Y}.

\begin{figure}[htp!]
	\centering
	\includegraphics[width=0.45\textwidth]{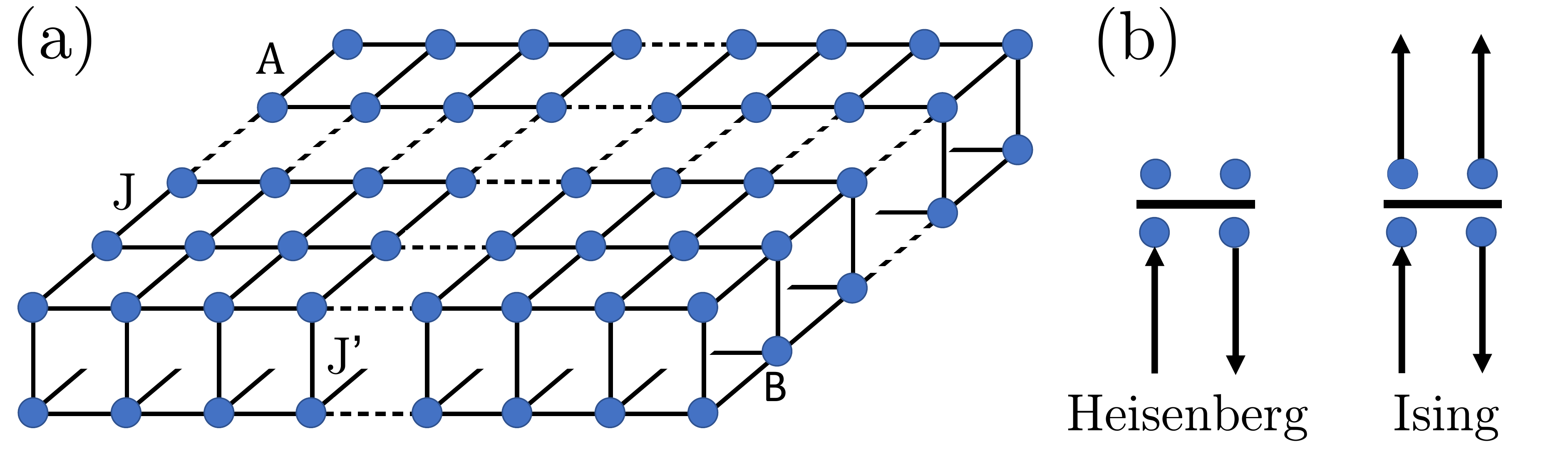}
	\caption{(a)  Spin-$1/2$ Bilayer model on an $L\times L$ square lattice. $L$ is the length of each layer. $J$ is the nearest-neighbor intra-layer ferromagnetic Ising interaction and $J'$ is the inter-layer antiferromagnetic Heisenberg interaction. (b) The schematic diagram of update lines on the space-time lattice, where the horizontal line represents the diagonal or off-diagonal operator, and above and below the line are four legs of a vertex. When the update line meets a Heisenberg operator, only the neighbor leg of the initial leg creates an update line (left panel). While all other three vertex legs create update-lines when it meets a Ising operator (right panel).}
	\label{fig:fig1}
\end{figure}

An interesting question can be asked: Can the EH, which is used to describe quantum entanglement, be a classical model (e.g., Ising model)? Intuitively, the two are contradictory. For example, all the eigenstates (as well as ground state) of Ising model are classical direct-product-state without any entanglement. However, EH itself is different from the original Hamiltonian, and we cannot exclude the possibility that this seemingly contradictory situation actually occurs. In fact, such an example has been successfully constructed in this article on a bilayer lattice, and the result is not limited to this case according to our discussion. Bilayer lattice can be divided into two 2D layer parts, which is well suited for studying the 2D entanglement properties. As the simplest layered structure, it has been revealed many novel physical phenomena in 2D systems~\cite{PhysRevLett.74.2303,PhysRevB.90.195131,PhysRevB.54.6640,PhysRevB.53.14481,Wang2006bilayer,Trambly2010,Trambly2012,Bistritzer_TBG,ROZHKOV20161,Santos2007,Santos2012,bernevig2020tbg3,YDLiao2019PRL,BBChen2020,ycwang2021,XuZhang2021,panDynamical2022,zhang2021superconductivity}.To answer above question, we design a bilayer model that each spin in the system (the first layer) is coupled to the environment (the second layer) with antiferromagnetic (AFM) Heisenberg interaction which introduces entanglement between two layers. This bilayer model with spin-$1/2$ is defined on an $L\times L$ square lattice with intra-layer ferromagnetic (FM) Ising interaction and inter-layer AFM Heisenberg interaction, as shown in Fig.~\ref{fig:fig1} (a).
The Hamiltonian is given by
\begin{equation}
    H=-J\sum_{\langle i,j\rangle}(S^z_{A,i}S^z_{A,j}+S^z_{B,i}S^z_{B,j})+J'\sum_{i}S_{A,i}S_{B,i},
    \label{eq:eq1}
\end{equation}
where $S_{A,i}$ and $S_{B,i}$ are spin operators at site $i$ of the $A$ (upper) and $B$ (bottom) layer, while $S^z$ represents the spin operator along $z$-axis. $\langle i, j \rangle$ denotes a pair of the nearest-neighbor sites on the same layer with periodic boundary condition. $J (>0)$ is the intra-layer Ising coupling strength and $J' (>0)$ is the inter-layer Heisenberg interaction.

In this paper, we develop a highly efficient cluster update scheme of stochastic series expansion (SSE) QMC algorithm (Sec.~\ref{method}) and investigate the ground state phase diagram of the spin-$1/2$ bilayer model in Sec.~\ref{pd}. The transition point and the critical exponents obtained from QMC simulations demonstrate that this continuous quantum phase transition (QPT) belongs to the 3D Ising universality. Then we measure the imaginary time correlations of the EH by a newly developed QMC method~\cite{2021arXiv211205886Y} to prove that the EH is always diagonal in the whole region of the phase diagram. We give an argument to explain how quantum model can emerge a classical EH in Sec.~\ref{eh}, and a conclusion in Sec.~\ref{cl}.

\section{Method}\label{method}
In this part, we will introduce our efficient update scheme within the SSE frame to deal with the Heisenberg-Ising mixed interactions. Readers who are not interested in the algorithm can skip to the next section, it doesn't affect the physical results. Although conventional directed loop SSE algorithm~\cite{Syljuaasen2002,Sandvik2010} can be employed to solve the model which contains both Ising and Heisenberg interactions, we found it works non-effectively due to the pure Ising interaction. Therefore, we have developed a new cluster method based on SSE combining the operator loop updates for Heisenberg model~\cite{Sandvik1991,Sandvik1999} and cluster updates for transverse-field Ising model~\cite{Sandvik2003}. Details of the algorithm are explained below. We note the readers who are not familiar with the SSE may need detailed SSE tutorial~\footnote{Prof. Sandvik has open source code and tutorial on his website \href{http://physics.bu.edu/~sandvik/programs/ssebasic/ssebasic.html}{http://physics.bu.edu/~sandvik/programs/ssebasic/ssebasic.html}} to learn more basics.

In the SSE QMC method, the partition function can be written as~\cite{Sandvik1991,Sandvik1999}
\begin{equation}
  \begin{split}
   Z &=\mathrm{Tr}\left\{e^{-\beta H}\right\}\\
   &=\sum_{\alpha}\sum_{S_{M}}(-1)^{n_{2}}\frac{\beta^{n}(M-n)!}{M!}\left\langle\alpha\left|\prod_{p=0}^{M-1}H_{a(p),b(p)}\right|\alpha\right\rangle,
 \end{split}
\end{equation}
where $\alpha$ sum over all states in the system, $S_{M}$ is the strings of the bond operators, and $M$ is the series expansion cut-off. $n$ is the total number of the operators and $n_2$ is the number of off-diagonal operators. $\beta$ is the inverse temperature.
Firstly, we add constant $\frac{1}{4}$ into the Hamiltonian to avoid the sign problem as we usually do in SSE~\cite{Sandvik1991,Sandvik1999,ZY2019,yan2020improved}, and separate it into the Ising part ($H_I$) and the Heisenberg part ($H_H$):
\begin{equation}\label{eq:3}
\begin{split}
    H&=-\sum_{\langle ij\rangle}H_{I}-\sum_{i}H_H,\\
    H_{I}&=J[(S^z_{A,i}S^z_{A,j}+\frac14)+(S^z_{B,i}S^z_{B,j}+\frac14)],\\
    H_{H}&=H_{1,H}+H_{2,H}\\
             &=J'[(\frac14-S^z_{A,i}S^z_{B,i})+\frac12(S^+_{A,i}S^-_{B,i}+S^-_{A,i}S^+_{B,i})],
\end{split}
\end{equation}
where $H_{I}$ is the diagonal operator as well as $H_{1,H}=J'(\frac14-S^z_{A,i}S^z_{B,i})$, and $H_{2,H}=\frac{J'}{2}(S^+_{A,i}S^-_{B,i}+S^-_{A,i}S^+_{B,i})$ is the off-diagonal operator. Taking $S^{z}$ as a complete set of basis for the system, the non-zero matrix elements are
\begin{equation}
\begin{split}
\langle \uparrow\uparrow|H_{I}|\uparrow\uparrow\rangle &= \langle \downarrow\downarrow|H_{I}|\downarrow\downarrow\rangle=\frac{J}2,\\
\langle \uparrow\downarrow|H_{1,H}|\uparrow\downarrow\rangle &= \langle \downarrow\uparrow|H_{1,H}|\downarrow\uparrow\rangle=\frac {J'} 2,\\
\langle \uparrow\downarrow|H_{2,H}|\downarrow\uparrow\rangle &=\langle \downarrow\uparrow|H_{2,H}|\uparrow\downarrow\rangle=\frac {J'} 2.
\end{split}
\end{equation}

The sampling procedures includes the diagonal updates and the cluster updates. The diagonal updates either insert or remove a diagonal operator between two states with acceptance probabilities, which are regulated by the detailed balanced condition. And the cluster updates change the types of operators by flipping all spins on the clusters with probability $1/2$.

We describe the updating scheme in the following steps:
\begin{enumerate}
	\item \textbf{Diagonal update}\\
	We go through the operator strings along the imaginary time direction and either remove or insert a diagonal operator on a randomly selected bond. If the chosen bond has two parallel (anti-parallel) spins on its endpoints, the insertion of Heisenberg (Ising) operator is forbidden. So, we will have several conditions:
	\begin{enumerate}
		\item For a diagonal operator ($H_{I}$ or $H_{1,H}$),
		we removed it with the acceptance probability, \begin{equation}
		P_{remove}=\min\left(\frac{M-n+1}{\beta (N_{H}J'+2N_{I}J)/2},1\right),
		\end{equation}
		where $N_{H}$ is the number of inter-layer bonds while $N_{I}$ is the number of intra-layer bonds in each layer, thus $N_{I}=2N_{H}$ according to our definition.
		\item For a null operator, we substitute it with a diagonal operator $H_{I}$ or $H_{1,H}$  by the procedures below.
		\begin{enumerate}
			\item Firstly, we decide which type of diagonal operators to insert. We choose to insert $H_{I}$ with probability \begin{equation}
			P(I)=\frac{2N_{I}J}{N_{H}J'+2N_{I}J},
			\end{equation} or $H_{H}$ with probability $P(H)=1-P(I)$.
			\item Once the decision has been made, we need to choose a certain position for the operator with probability $1/N_{H}$ or $1/2N_{I}$. Then the operator will be inserted with the acceptance probability
			\begin{equation}
			P_{insert}=\min\left(\frac{\beta(N_{H}J'+2N_{I}J)/2}{M-n},1\right).
			\end{equation}
			
		\end{enumerate}
	\item For an off-diagonal operator, we ignore it and go to the next slice of the imaginary time.
	\end{enumerate}
	\item \textbf{Cluster update}
	\begin{enumerate}
		\item The schematic diagram of update lines in the configuration space is shown in Fig.~\ref{fig:fig1} (b). QMC is based on the path integral in imaginary time~\cite{Syljuaasen2002,Sandvik2010AIP,zyan2018}. The evolution of the system can be mapped on the space-time lattice. For each step of the imaginary time propagation, an operator act on the initial state and a new propagated state is obtained. Fig.~\ref{fig:fig1} (b) shows the action of the operator, where the horizontal line represents the diagonal or off-diagonal operator, and above and below the line are four legs of a vertex. When an update line meets an Ising operator (intra-layer), all the other three vertex legs create update lines. While only the neighbor leg of the initial leg creates an update line when it meets a Heisenberg operator (inter-layer).
~\\
		
		For the off-diagonal updates, we follow two rules to construct the clusters: (1) the two neighbour legs on the same side of a Heisenberg operator $H_{H}$ belong to the same cluster; and (2) the four legs of a Ising operator
		$H_{I}$ belong to one cluster.
        By this way, The update line evolves and forms a cluster eventually.
		\item Then the clusters constructed from above rules are flipped with probability $1/2$, which is according to the Swendsen-Wang cluster updating scheme.
	\end{enumerate}
	
\end{enumerate}
All the details about the update accept-probability is similar to the previous SSE: the update of Heisenberg operators obeys the operator loop method for Heisenberg model~\cite{Sandvik1991,Sandvik1999} and that of Ising operators obeys the cluster updates for transverse-field Ising model~\cite{Sandvik2003,zhou2020quantum,zhou2020quantumstring,ZZrk2021}.

\section{Phase diagram and critical exponents}\label{pd}
\begin{figure}[htp!]
	\centering
	\includegraphics[width=0.45\textwidth]{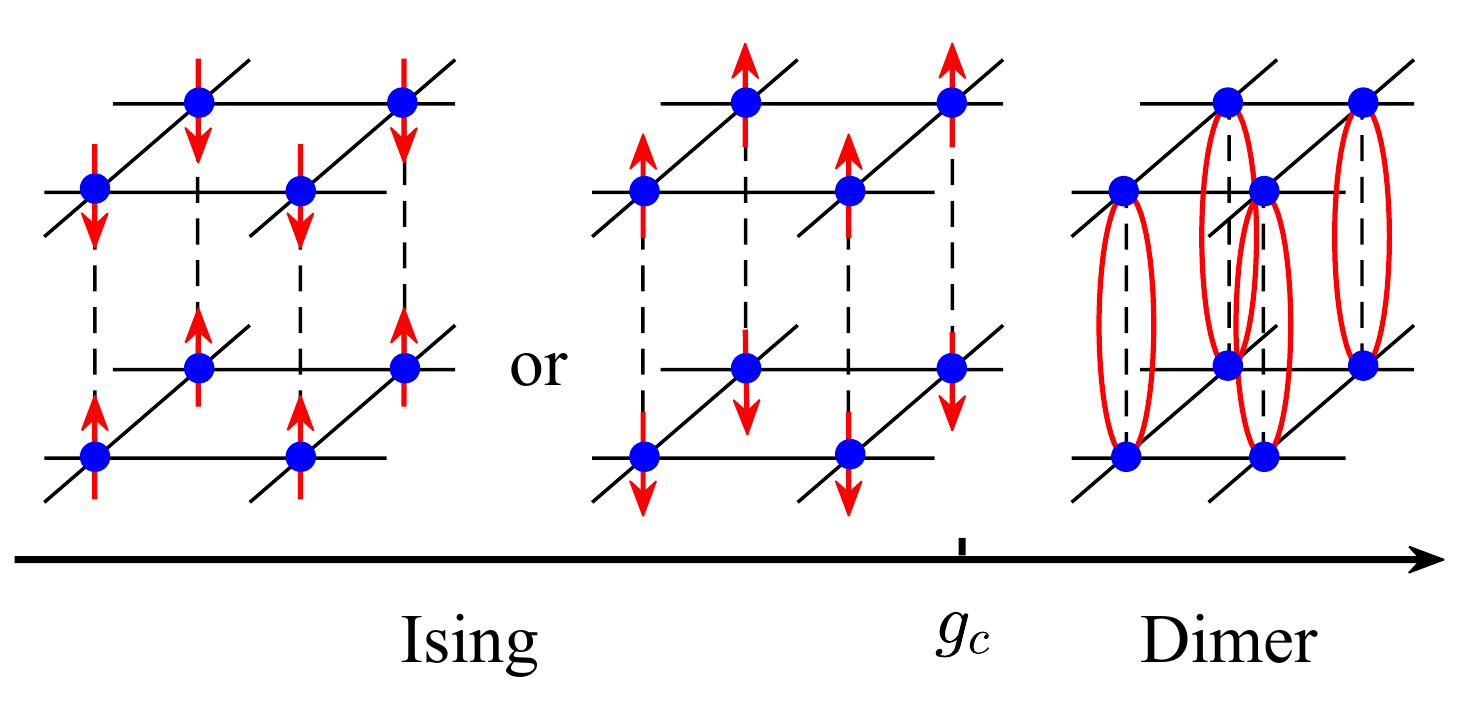}
	\caption{The ground state phase diagram of the square lattice bilayer model. When $g>g_c$, there is a two-fold degenerate FM phase, where spins in the same layer are aligned parallel. For $g>g_c$, the ground state corresponds to a dimerized phase where spins in different layers coupling into singlets.}
	\label{fig:phasediagram}
\end{figure}

Using the QMC method we developed for the spin-$1/2$ bilayer model, we study the ground states and the critical properties of the system. Here we set the coupling ratio $g=J'/J$. When $g \rightarrow 0$, the intra-layer FM Ising coupling $J$ dominates, the system favors a FM phase where all the spins in the same layer are parallel. And when $g \rightarrow \infty$, which means the inter-layer AFM Heisenberg interaction $J'$ is very large, the state corresponds to a dimerized phase where the two spins of different layers are coupled into singlet. The phase diagram of the system is shown in Fig.~\ref{fig:phasediagram}.

To further understand the phase transition in this system, we calculated several physical observables, including the order parameter $m$, the Binder cumulant $U_2$, the susceptibility $\chi$, and the spin-spin correlation function $G(r)$. The order parameter $m$ is defined as the magnetization difference of upper and bottom layer
\begin{equation}
    m=\sum_{i}(S^z_{A,i}-S^z_{B,i}).
\end{equation}
The Binder cumulant\cite{Binder.1981,binder1981finite}  $U_2$ is given by
\begin{equation}
U_2=\frac{3}{2}\left(1-\frac{1}{3}R_2\right),
\end{equation}
where $R_2= \left\langle m^4\right\rangle/\left\langle m^2\right\rangle^2$ is the binder ratio \cite{Binder.1981,Binder.1984} of the order parameter $m$. The susceptibility $\chi$ can be expressed as
\begin{equation}
\chi=\frac{\beta}{N}\left(\left\langle m^2\right\rangle-\left\langle |m|\right\rangle^2\right),
\end{equation}
where $N$ is the total number of spins. The spin-spin correlation function $G(r)$ can be written by
\begin{equation}
G(r) = \left\langle S_0^zS_r^z\right\rangle,
\end{equation}
where $r$ is the distance between two sites.

\begin{figure}[htp!]
	\centering
	\includegraphics[width=0.45\textwidth]{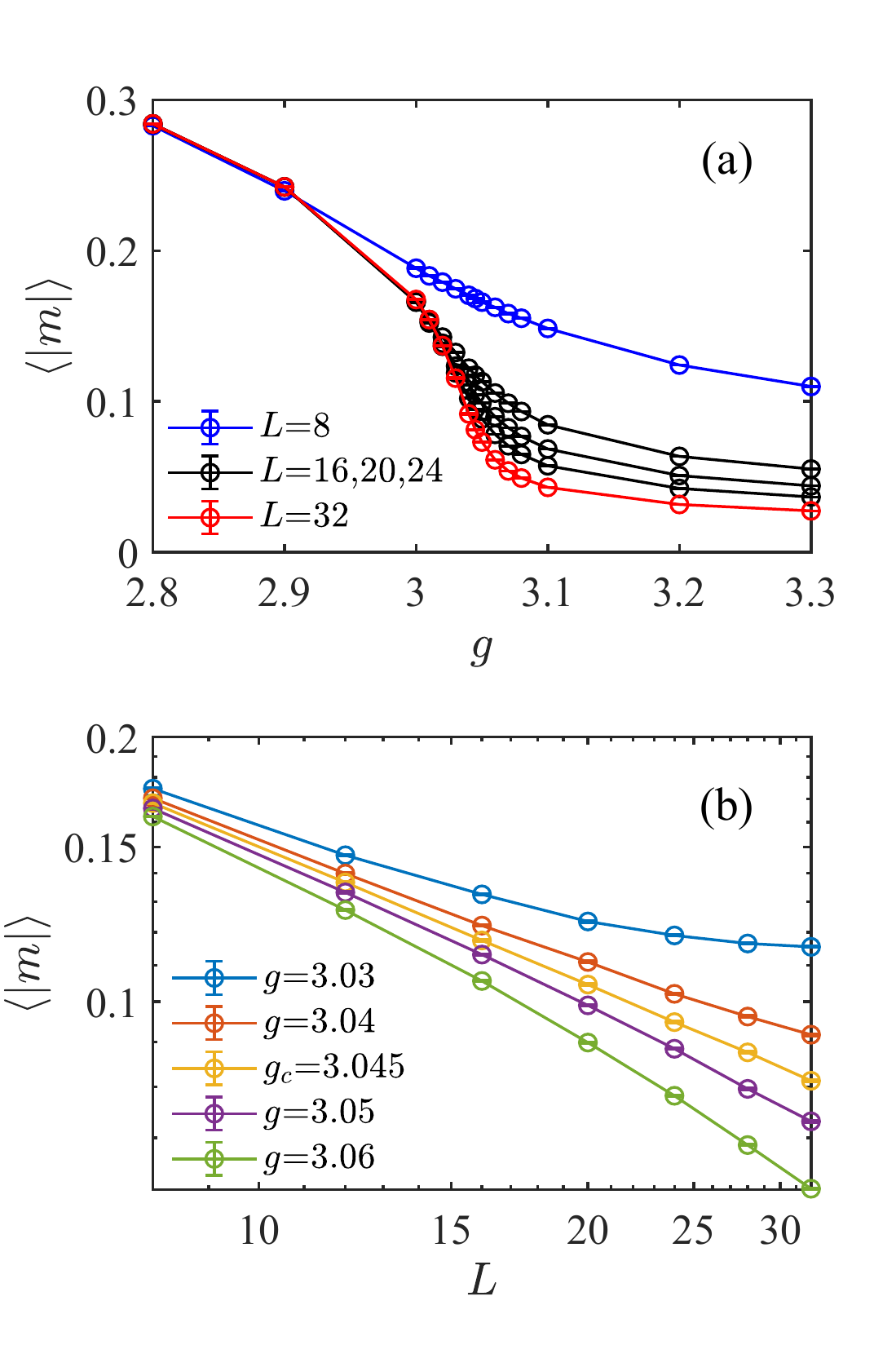}
	\caption{(a) The order parameter $m$ as a function of the coupling ratio $g$. As $g$ increases, $\langle |m| \rangle$ eventually decay to zero indicating the vanishing of the FM phase. We calculate data with smaller interval nearby the critical point to illustrate the continuous quantum phase transition in the system. (b) The log-log plot of $\langle|m|\rangle$ around the transition point $g_c$. We find in the FM phase the order parameter decays to a constant while to 0 in the dimerized phase. And at the critical point, it shows a power-law decay. }
	\label{fig:m}
\end{figure}

The order parameter $m$ versus $g$ for different system sizes is shown in Fig~\ref{fig:m}(a), which depicts the vanishing of the FM phase. As $g$ increases, $\langle |m| \rangle$ tends to be zero, where two spins in different layers form a singlet. To study the phase transition between the FM state and the dimerized phase, we show a log-log plot of $\langle|m|\rangle$ in Fig.~\ref{fig:m}(b) to analyse its different decay behaviors in different phases. We found in the FM phase ($g<g_c$), $\langle |m| \rangle$ drops to a constant as increasing the system size. While in the dimerized phase ($g >g_c$), the order parameter exponentially decays to 0 as $L\rightarrow\infty$. At the critical point $g_c$, $\langle |m| \rangle$ shows a nontrival power-law decay.

\begin{figure}[htp!]
	\centering
	\includegraphics[width=0.45\textwidth]{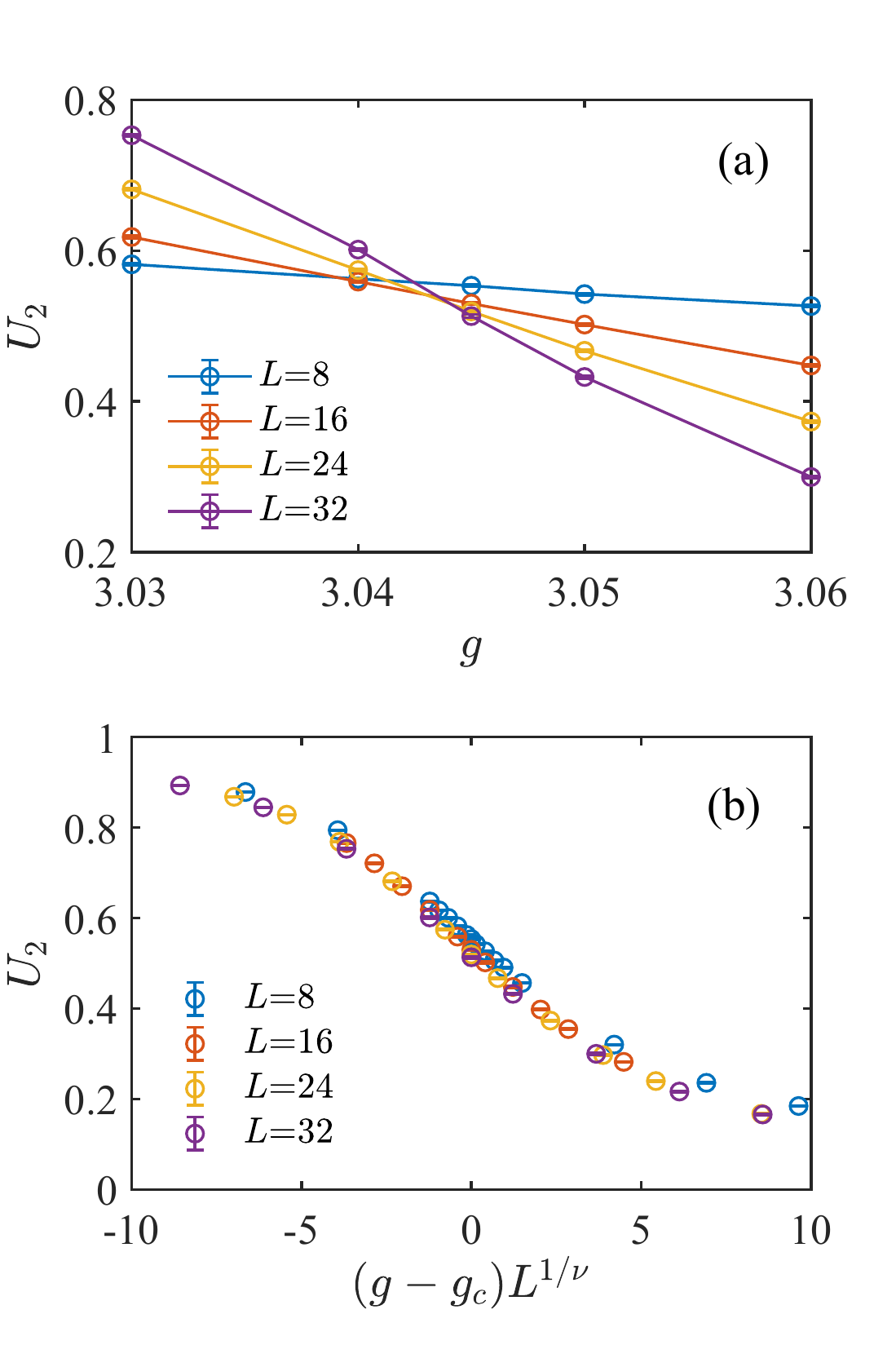}
	\caption{(a) The binder cumulant $U_2$ versus $g$. We determine the critical point $g_c$ by extrapolating the crossing points of $U_2$ curves with different system sizes, which gives $g_c=3.045(2)$. (b) The high-quality data collapse of the binder cumulant at the critical point. Here we use the 3D Ising critical exponent $\nu=0.63$ in the scaling, which suggests our numerical results satisfies the 3D Ising universality class.}
	\label{fig:binder}
\end{figure}

To determine the critical point $g_c$, we obtain the Binder cumulant and its data collapse with 3D Ising critical exponents, which are shown in Fig.~\ref{fig:binder}. We calculate systems with $L=8, 16, 24, 32$. The drift of the crossing points of different system sizes in Fig.~\ref{fig:binder}(a) indicates the finite-size effect in the system. So we perform the extrapolation of these crossing points to obtain the critical point in the thermodynamic limit, which gives $g_c=3.045(2)$. The high-quality data collapse of $U_2$ in Fig.~\ref{fig:binder}(b), where $\nu=0.63$ is the 3D Ising critical exponent\cite{Pelissetto2002, Kleinert.1999}, suggests that the FM phase to dimerized phase transition belongs to the 3D Ising universality class.

\begin{figure}[htp!]
	\centering
	\includegraphics[width=0.45\textwidth]{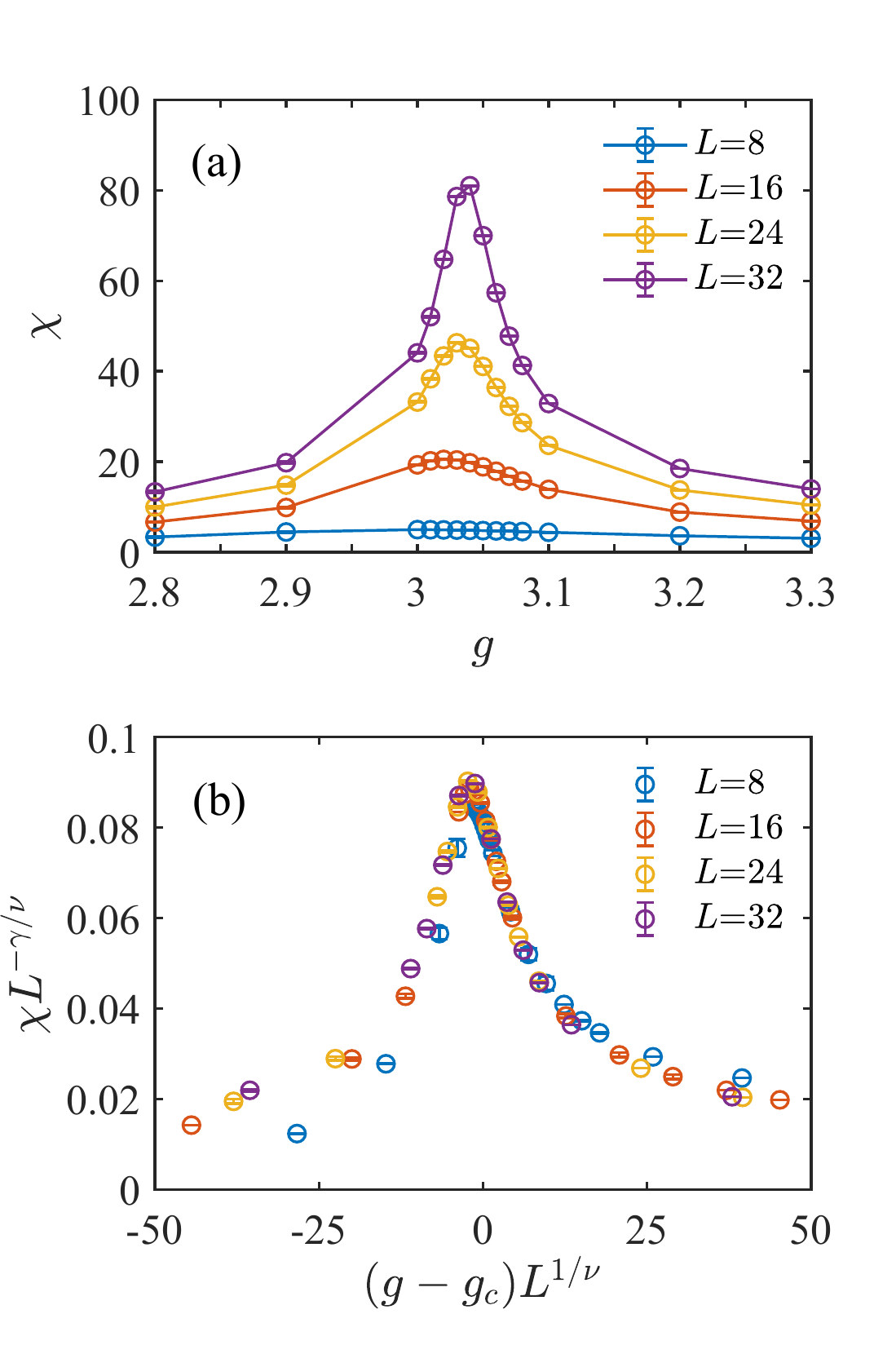}
	\caption{(a) The susceptibility $\chi$ as a function of $g$. The peak in the critical point diverges as increasing the system size $L$. (b) The data collapse of the susceptibility is observed with 3D Ising exponents $\nu=0.63$ and $\gamma=1.24$, and our suspection for the universality class is confirmed again.}
	\label{fig:chi}
\end{figure}

Besides, the susceptibility $\chi$ illustrated in the Fig.~\ref{fig:chi} shows a peak in the transition point, which is supposed to diverge with increasing system size $L$. Data with smaller interval is calculated to show the peak more clearly. We also perform the data collapse according to the scaling law $\tilde{\chi}(L,t) = L^{-\gamma/\nu} {\chi} (tL^{1/\nu})$\cite{Fisher.1972}. Here we use the Ising critical exponent $\nu$ and $\gamma$, where $\gamma=1.24$. The data collapse is very good near the critical point.


\begin{figure}[htp!]
	\centering
	\includegraphics[width=0.45\textwidth]{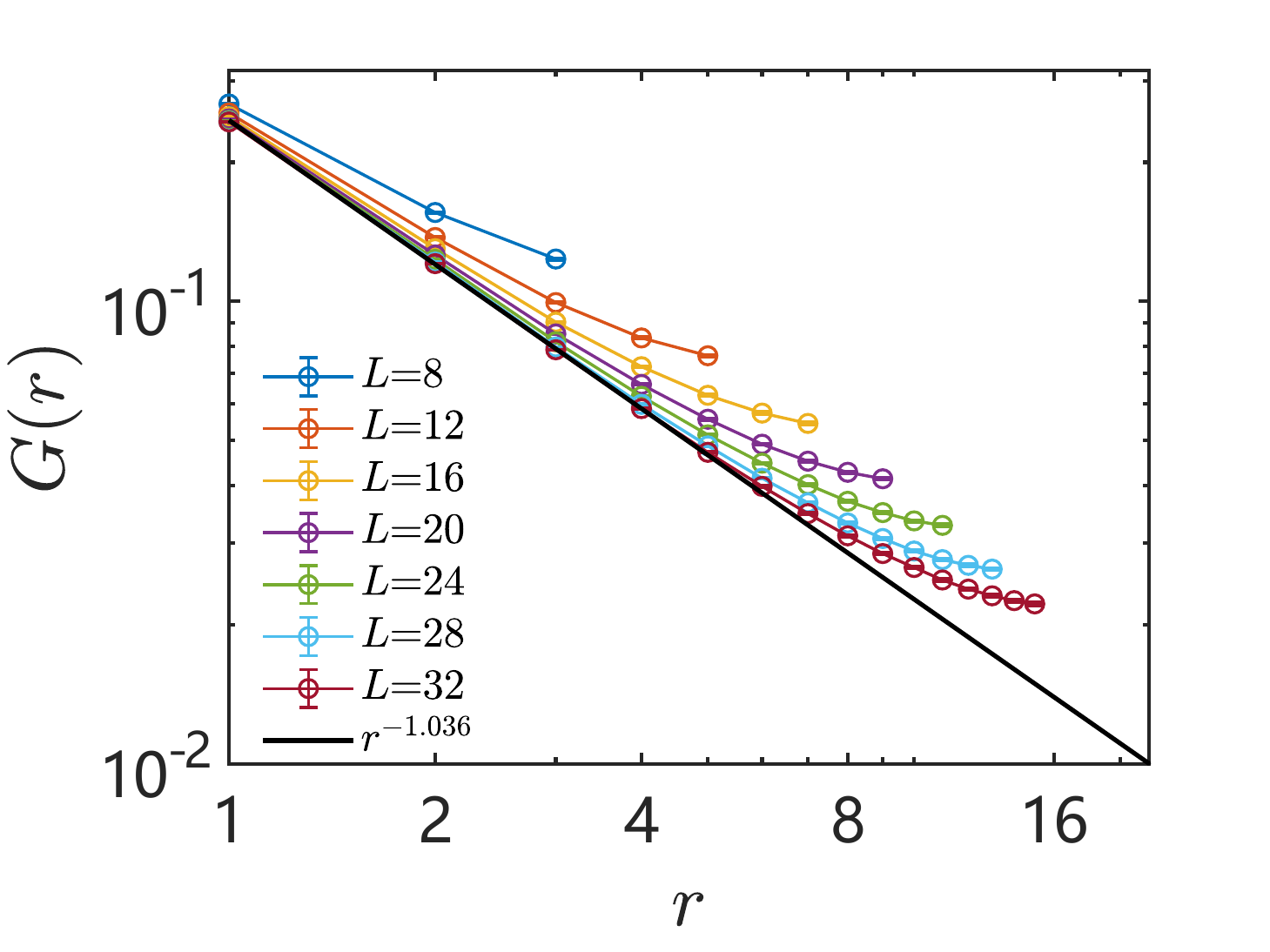}
	\caption{ The log-log
		plot of the equal-time spin-spin correlation function $G(r)$ as versus the distance
		$r$ at $g_c=3.045(2)$, which fits the scaling relation $G(r)\sim r^{-(1+\eta)}= r^{-1.036}$. Here we also use the 3D Ising critical exponent $\eta = 0.036$, which corresponds to the black reference line. The scaling of the  $G(r)$ shows a power-law decay at the transition point. }
	\label{fig:G}
\end{figure}

We also consider the equal-time spin-spin correlation function $G(r)$ of a single layer, which is shown in Fig.~\ref{fig:G} in a log-log plot. At the critical point, $G(r)$ should decay as a power-law $G(r) \sim r^{-(1+\eta)}$. As a comparison, we plot the $\eta = 0.036$ line of the 3D Ising universality. These numerical evidences confirm that there is a 3D Ising transition between the FM phase and the dimerized phase.

\section{Entanglement Hamiltonian}\label{eh}
The Heisenberg interactions favor to bound the spins of different layers into singlets and induce the quantum entanglement between two layers. Obviously, the strength of the quantum entanglement between layer $A$ and layer $B$ should be different in FM and dimerized phases. We try to extract the information of the EH from QMC simulation through recently developed method which gained the ES successfully~\cite{2021arXiv211205886Y}. According to this method, we can obtain the imaginary time correlations of the entanglement Hamiltonian $H_A=-{\rm ln}(\rho_A)$ via constructing the replica manifold to simulate ${\rm Tr}(\rho_A^n)$. Here $\rho_A={\rm Tr}_B(\rho)$ is the RDM of layer $A$, and $\rho$ is the density matrix of the bilayer system. The key point of this method is that the imaginary time correlation function can be obtained by simulating ${\rm Tr}(\rho_A^n)$ via QMC~\cite{2021arXiv211205886Y,Assaad2014,Assaad2015,Parisen2018}. Therefore, if we treat the $n$ as the inverse temperature $\beta$ for the entanglement hamiltonian $H_A$, every $\rho_A=e^{-H_A}$ can be seen as an imaginary time evolution operator with the time length $1$.
\begin{figure}[htp!]
	\centering
	\includegraphics[width=0.45\textwidth]{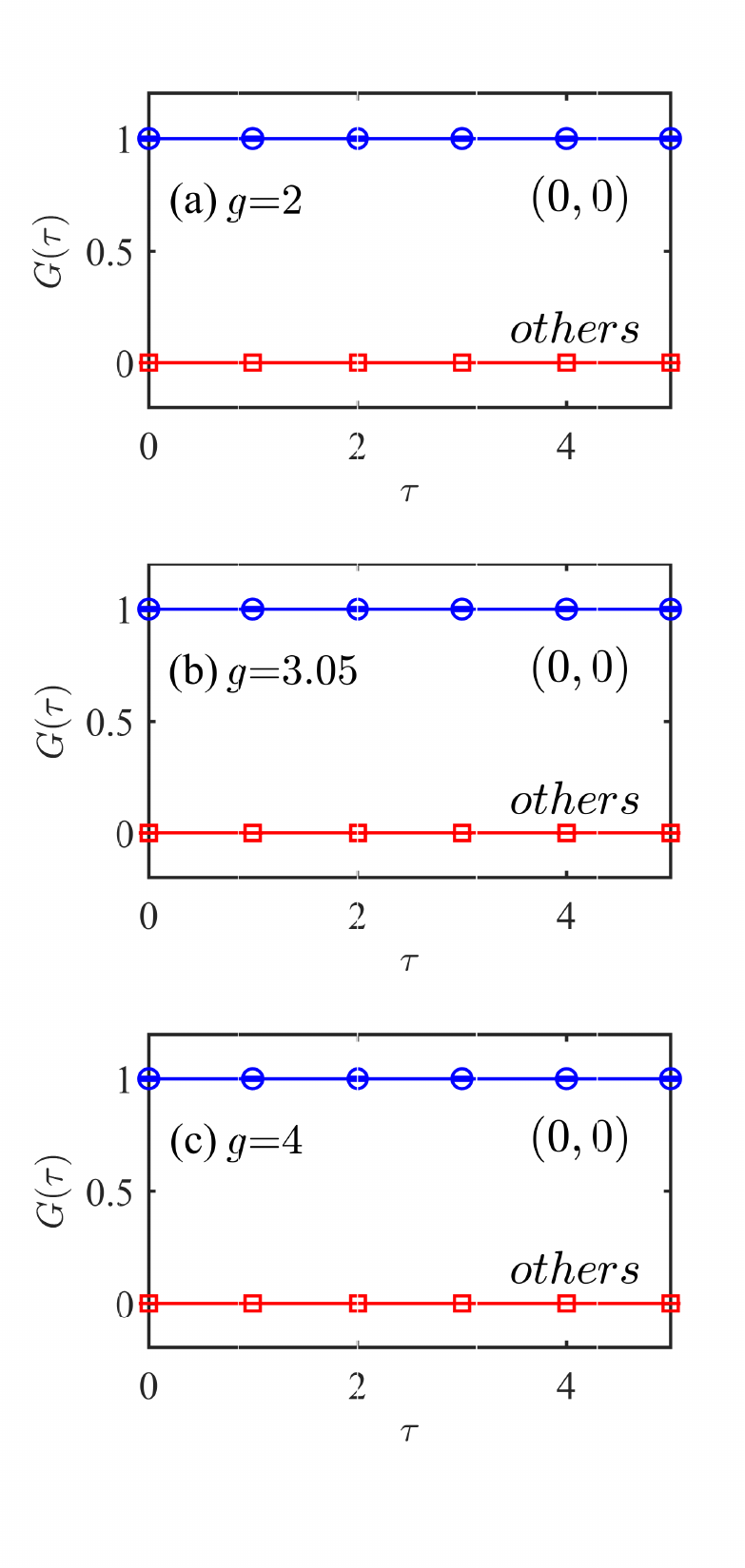}
	\caption{$G(\tau)$ of entanglement Hamiltonian in the momentum space of (a) the FM Ising phase, (b) the critical point, and (c) the dimerized phase. $G(\tau)$ in the $\Gamma$ point (blue) and other momentum points (red) are calculated with different $g$. We find $G(\tau)$ never decay in any phase, which is equal to $1$ only at $(0,0)$ point and $0$ in other momentum points.}
	\label{fig:Gg}
\end{figure}

We now discuss the $S^z$ imaginary time correlations of the EH, i.e., $G(\tau)=\langle S^z(\tau)S^z(0) \rangle_{EH}$. It is interesting that we find $G(\tau)$ doesn't decay regardless the phase of the system. As is shown in Fig.\ref{fig:Gg}, the correlation function is always a constant in FM Ising phase, dimer phase, or at the phase transition point. Its value is equal to $1$ (blue line) only at $k=(0,0)$ momentum point~\footnote{Because the A part is still periodic boundary condition even if tracing out the B part, we can well define the momentum basis in this case. We can rewrite the Hamiltonian from the real space into momentum space through a Fourier transform. In fact, it means we use the momentum basis $|S^z_k \rangle$ to instead of real space basis $|S^z_i \rangle$. The Fig.7 is the imaginary time correlation function of momentum space which can be obtain from real space correlations via Fourier transform, that is, $C(k)=\sum_r e^{-kr}C(r)$. Because here the A is 2D system, the $k$ and $r$ are also 2D, $r=(i,j)$ with $i,~j=0,~1,~..,~L-1$ and $k=2\pi/L(m,~n)$ with  $m,~n=0,~1,~..,~L-1$}, and $0$ anywhere else (red line). In the real space, all the imaginary time correlation $G_i(\tau)=\langle S_i^z(\tau)S_i^z(0) \rangle_{EH}$ of site $i$ is always $A$. All these evidences demonstrate that the EH must be a diagonal matrix in $S^z$ basis, which means it is indeed a classical Ising model without any quantum fluctuation. It is definitely a non-trivial phenomena that such a classical model emerges in the quantum entanglement.

\begin{figure}[htp!]
	\centering
	\includegraphics[width=0.4\textwidth]{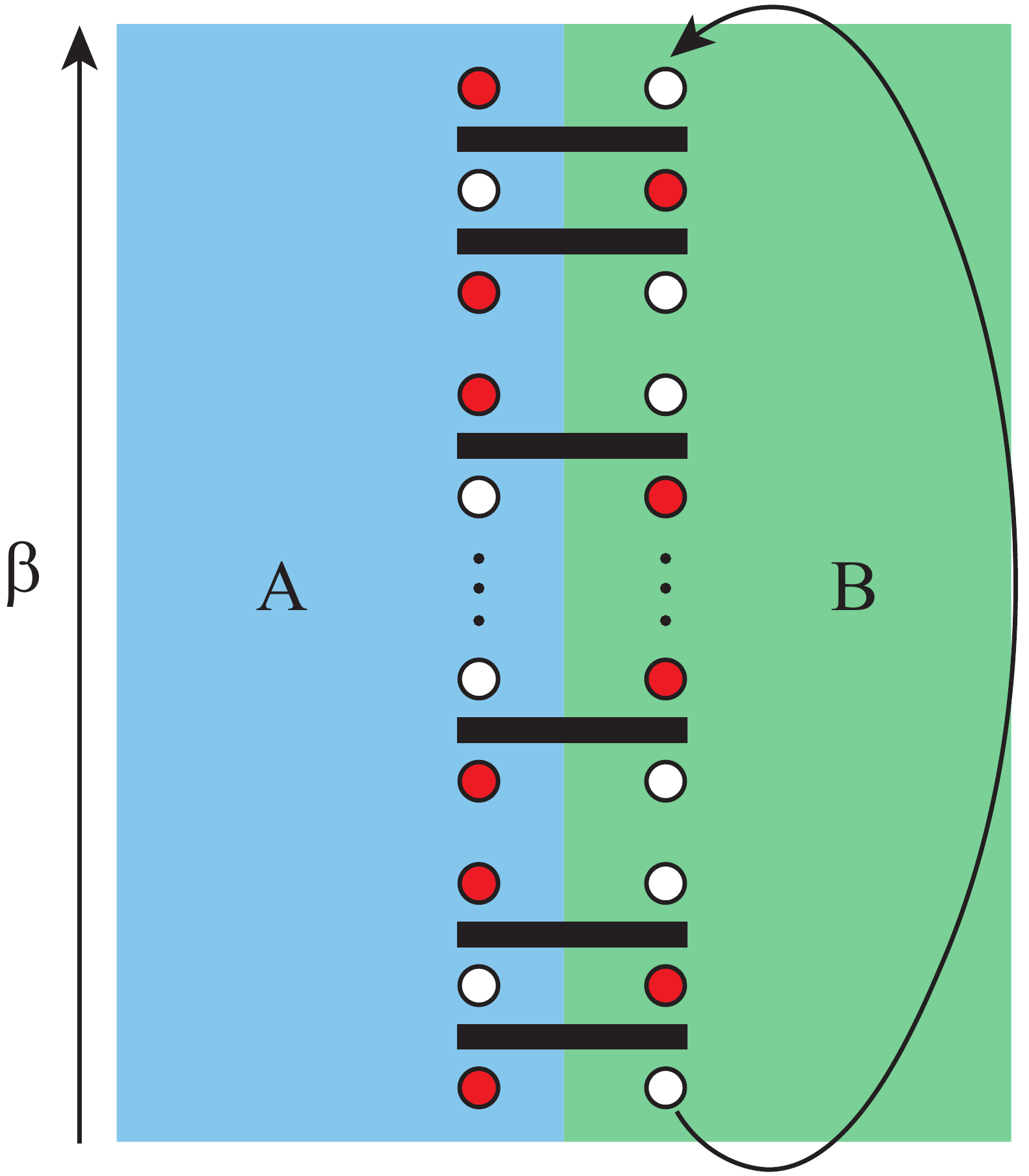}
	\caption{The schematic diagram of the path-integral of the RDM $\rho_A$ in the SSE framework. The imaginary time boundary of B part is periodic because of the $\mathrm{Tr}_B(\rho)$. The red(white) circle corresponds to spin up(down). The horizontal black bars depict
off-diagonal operator an off-diagonal operator. $A$($B$) refers to the layer $A$($B$). For convenience, we only draw one spin to represent one layer.}
	\label{fig:QMC}
\end{figure}

Back to the effective Hamiltonian of a single layer, it seems like an Ising Hamiltonian with quantum fluctuation term $S^x_i$ on every site $i$. As shown in Fig.\ref{fig:QMC}, a configuration of the path-integral inside the RDM $\rho_A$ contains several off-diagonal operators of Heisenberg interaction which flip the spins along the time evolution. Because the $(S_{A,i}^+S_{B,i}^-+S_{A,i}^-S_{B,i}^+)$ acts on the inter-layer, it takes an effective operator as $(S_i^++S_i^-)\sim S_i^x$ on position $i$ for a single layer $A$. Therefore, the effective Hamiltonian of one layer has quantum fluctuations, which truly different from the single layer EH obtained from the above imaginary time correlations. In addition, there is a qualitative analogy: when the Heisenberg interaction is strong, the effective $S^x_i$ term on
subsystem $A$ also becomes strong. The dimerized phase of total system can be seen as a paramagnetic phase of subsystem $A$ in this sense.

Finally, we find the explanation from the path integral of the RDM. As the EH is diagonal, it could be inferred that the RDM $\rho_A$ is diagonal, we can focus on proving $\rho_A={\rm Tr}_B(\rho)$ is diagonal. In our bilayer case, all the off-diagonal operators are from the Heisenberg interactions which act on the inter-layer bonds only. To obtain $\rho_A$, we have to trace layer $B$, which is regarded as the environment. Trace operator requires the boundary condition of imaginary time to be periodic, which means the off-diagonal operators should act even times in layer $B$ to keep the upper/lower spin configurations same, as shown in Fig.\ref{fig:QMC}. Therefore, the off-diagonal operators must also act on layer $A$ even times to ensure the RDM $\rho_A$ is diagonal, that is, the upper/lower spin configurations of $A$ should be same as well.

Perhaps the readers think the above explanation is very tricky and dependent on the detailed QMC method. In fact, it is just based on a path integral frame generally. $\rho_A =\mathrm{Tr}_{B}\left\{e^{-\beta H}\right\}$ is the definition of the RDM, and it can be rewritten via Taylor series expansion as $\mathrm{Tr}_{B}\left\{a_0+a_1H+a_2H^2+...+a_nH^n\right\}$. The trace $B$ requires the initial and final configurations (bra and ket) of $B$ should be same. It means the off-diagonal $H$ operators should not change the initial configuration of $B$ odd times. In our case, all the off-diagonal operators act on both $A$ and $B$, that is, the number of off-diagonal operators must be even. Because the off-diagonal operators work as $S_A^+S_{B}^-+h.c.$, it means the initial configurations of $A$ and $B$ should be flipped same times. Therefore, the $A$ configurations also keep unchanged. Furthermore, this conclusion can be generalized to more normal cases: all the interactions in intra system and intra environment are diagonal while the off-diagonal interactions only exist in entangled boundary region.

Although the EH is always classical in these conditions, the EH levels become higher while the coupling grows up. Because the EE defined as $S_A=-{\rm Tr}(\rho_A {\rm ln} \rho_A)$ increases as the eigen-levels of EH, the classical EH is consistent with the strong EE. In fact, for a real direct product state without entanglement, its RDM is still a pure state and EH only has one level, which means there is not any entanglement.

\section{Conclusion}\label{cl}
In this paper, we study a bilayer spin-$1/2$ model with intra-layer FM Ising interaction and inter-layer antiferromagnetic Heisenberg interaction. An efficient cluster update of SSE is developed to overcome the bounce problem~\footnote{When we use the directed loop update to sample the QMC configurations, if the $S^zS^z$ term is much larger than $S^+S^-+h.c.$ term, the update lines will always go back (bounce) when meet vertex. It makes the samplings much ineffective. Thus, strong Ising interactions usually lead to hard simulation for QMC} in directed loop update in pure Ising interactions. Using this developed QMC method, we determine the critical point $g_c=3.045(2)$ for the phase transition between the FM phase and the dimerized phase. Moreover, the analyses of the critical exponents indicates that the phase transition belongs to the 3D Ising universality class.

Furthermore, we find the quantum entanglement Hamiltonian of single layer is a pure classical Ising model without any quantum fluctuation, although its effective Hamiltonian seems like a transverse-field Ising model. Therefore, we give a more general conclusion for the reason why a classical entanglement Hamiltonian can emerge. By calculating the imaginary time correlations of the EH, we find the EH is diagonal in both FM phase and dimerized phase, which can be explained by the path integral of the RDM.

\section*{Acknowledgements}
We thank Chang-Qin Wu, Bin-Bin Chen, and Zi Yang Meng for helpful discussions. We acknowledge the support from Zhejiang Provincial Natural Science Foundation of China under Grant No. LZ23A040003. Q.-F.L. acknowledges financial support of the MERIT-WINGS course provided by the University of Tokyo, and the Fellowship for Integrated Materials Science and Career Development provided by the Japan Science and Technology Agency. Y.-C.W. and S.W. thank Beihang Hangzhou Innovation Institute Yuhang for support. The numerical calculations in this paper have received support from the supercomputing system at the High-Performance Computing Centre of Beihang Hangzhou Innovation Institute Yuhang. S.W. and B.-B.M. acknowledge
Beijng Paratera Tech(\url{https://www.paratera.com/}) for providing high-performance
computing resources that have contributed to the research results reported in this paper.

\bibliographystyle{apsrev4-1}
\bibliography{bilayer-ising}
\end{document}